\documentclass[5p]{elsarticle}

\usepackage{amssymb}

\usepackage{subfigure}

\journal{TIPP09 Proceedings in NIMA}

\begin{document}

\begin{frontmatter}



\title{The MEG Spectrometer at PSI}

%
%
%
%
%
\author{Paolo Walter Cattaneo on behalf of the MEG collaboration } 
\ead{Paolo.Cattaneo@pv.infn.it}
%
%
%
\address {INFN Pavia, Via Bassi 6, Pavia, I-27100, Italy }

%
%
%
%
%
\begin{abstract}

The MEG experiment is designed to search the Lepton Flavor Violating process
$\mu \rightarrow e^+\gamma$ \cite{mori-1999,meg2007-baldini}.
This search requires a high intensity muon beam stopping in a thin target 
with the maximum rate compatible with the background from combinatorial events.

The events are analyzed by a high resolution and fast liquid xenon
calorimeter and by a spectrometer composed by an array of ultra-light drift chambers
(DCH) for momentum measurement and a double layered timing counter (TC) for measuring
the $e^+$ time.

The design parameters and performance of the spectrometer during the
2008 physics run are described.

\end{abstract}

%
%
%
%
%

\end{frontmatter}


%
%
%
%
%
%

\section{Design parameters of the experiment}

The goal of MEG is improving the limit on $BR(\mu^+ \rightarrow e^+\gamma) = 
1.2\times 10^{-11}$ \cite{pdf2006} by two order of magnitude 
to probe new physics beyond the Standard Model.

The signal has a simple topology: monochromatic $e^+$ and $\gamma$ with $p_\gamma=
p_{e^+}=m_\mu/2$ moving in opposite directions ($\cos \Theta_{e^+\gamma} = -1$),
originating from the same space-time point.

The physical background comes from the radiative decay $\mu^+\rightarrow e^+\gamma\nu\nu$
with vanishing $\nu$ energy. The accidental background comes from a $e^+$ from $\mu^+$
Michel decay and a $\gamma$ with signal-like kinematic. The $\gamma$ may originate
from $\mu^+$ radiative decay, from $e^+$ Bremsstrahlung or from $e^+$ annihilation

In MEG the accidental background dominates. It is proportional to the instantaneous 
$\mu^+$ rate $R_\mu$, because of coincidence within the time resolution window. 
The optimal beam configuration is a DC beam with $R_\mu$ dependent on the detector resolution 
and background.

The experiment is located at the Paul Scherrer Institut near Z\"urich, Switzerland that
provides the most intense DC $\mu^+$ beam in the world.

The $\gamma$ detection is based on homogeneous liquid xenon
calorimeter. The design parameters and performances are discussed in \cite{ryu-tsukuba}.

The $e^+$ spectrometer is designed with an acceptance $\epsilon_{e^+} \approx 50\%$ and the
following FWHM resolutions: time $\Delta t_{e^+} = 100\,ps$, angular $\Delta \Theta_{e^+} =
10.5\,mrad$ momentum $\frac{\Delta p_{e^+}}{p_{e^+}} = 0.8\,\% $ and vertex position on target
$\Delta v_{e^+} = 2.1\,mm$.

\section{The $\mu^+$ beam}

The $\mu^+$ beam is obtained starting from a $\pi^+$ stopped beam. The $\mu^+$ from $\pi^+$ decaying 
close to the target surface (surface muons) are monochromatic with momentum 
$p_{\mu^+} = 27.8\mathrm{Mev/c}$.
They are transported through a Beam Transport Solenoid (BTS) into the MEG detector.

The beam intensity is steerable up to a maximum above $R_\mu=1.1\times 10^8\,\mu/s$. To cope with 
combinatorial background we use the intensity $R_\mu=3.0\times 10^7\,\mu/s$.

\section{The $e^+$ spectrometer}

The approach of the MEG spectrometer in satisfying the above design parameters in presence
of high $e^+$ rate consists in developing a particular configuration of the magnetic field
to sweep out quickly $e^+$ preventing excessive hit crowding in DCH and in minimizing 
material on $e^+$ path to reduce multiple scattering. The whole spectrometer is embedded in He gas.
A display of the spectrometer with a signal event is shown in Fig\ref{fig12}a.

\subsection{The COBRA magnet}

The COnstant Bending RAdius magnet is a thin superconducting magnet with an axial gradient
and a central value $B=1.26\,T$.
This configuration has two advantages: the maximum radius of a $e^+$ trajectory depends mainly on 
its total and not on its transverse momentum, therefore almost independent from its
polar angle, and the $e^+$ emitted at normal polar angle are swept away in a few turns.

\subsection{The target}

The target is made of polyethylene $205\,\mu \mathrm{m}$ thick with an ellipsoidal shape.
It is located at COBRA center tilted at $20.5^\circ$ with the beam axis.
Holes with $0.5\,\mathrm{cm}$ diameter are drilled in the target for easing alignment.

This configuration guarantees that $e^+$ emitted in the spectrometer acceptance
suffer low multiple scattering while the fraction of $\mu^+$ not stopping in the target
is minimized: in fact $\approx 70\,\%$ of the $\mu^+$ entering the BTS decay in the target.

\subsection{The Drift Chamber}

The $e^+$ momentum is measured with a set of 16 low mass Drift Chambers.
They are positioned at fixed $\phi$ with wires along z (beam direction).
Design and construction details are presented in \cite{malte-tsukuba}.

The chamber measures the $e^+$ momentum and kinematic parameters.
The measured tracks are extrapolated to match TC hit where their timing is precisely
measured and to the target to define the event vertex.

Problems with HV trips reduced significantly the DCH efficiency during the 2008 run
and worsened the momentum resolution compared to the design values. Chamber are being 
redesigned to remove this problem.

\subsection{The Timing Counter }

The timing counter (TC) is designed to provide trigger information on the momentum, relative
direction and time of the $e^+$. It is divided in two sectors, upstream and downstream the
target. Each sector consists of an inner layer made of $0.5\,\mathrm{cm}$ thick scintillating 
fibers read by APD and an outer layer made of $4\,\mathrm{cm}$ thick scintillating bars 
read by PMT insensitive to the magnetic field along their axis. 

The PMT output is sampled at 2GHz with a custom designed chip \cite{drs2007,drs2008} for offline 
timing reconstruction. The same chip is used for sampling the output of the liquid xenon calorimeter
that measured the $\gamma$ time.

The inner layer measures the $z$ coordinate while the outer one the $\phi$ coordinate and the time. 
The TC timing resolution is obtained studying events with multiple bar hits. The result of
$\sigma(t_{e^+}) = 54\,ps$ is very close to the design value $\sigma(t_{e^+}) = 45\,ps$.

\subsection{The Time Calibrations }

The TC tight timing requirements need a time intercalibration between bars and with the calorimeter. 
$t_{e^+}$ is measured on the bars and extrapolated to the vertex on the target using the track parameters 
measured with the DCH. $t_\gamma$ is measured on the inner surface of the calorimeter and 
extrapolated to the target. $t_{e^+}$ and $t_\gamma$ extrapolated to the target are required 
to match within the experimental resolution.

Each PMT on each bar sides provides a separate and independent measurement of time, related to the $e^+$
impact time on the n-th bar $t_{e^+}$ from

\begin{eqnarray}
T_{PMT0}^n = t_{e^+} + \frac{z}{v_{eff}} + b^n_0 \nonumber \\
T_{PMT1}^n = t_{e^+} + \frac{L-z}{v_{eff}} +b^n_1 \nonumber 
\end{eqnarray}
 
where PMT0 is on the inner side and PMT1 on the outer one. $z$ is the coordinate along the bar, 
$v_{eff}$ the effective velocity and $b^n_{01}$ the PMT offsets.

For cosmic rays, that hit the bars uniformly in $z$, the distribution of $\Delta T^n = T_{PMT1}^n - 
T_{PMT1}^n$ is centered at $b^n_1 - b^n_0$, so that $b^n_1$ are measured.

Besides Michel $e^+$ and cosmic rays, that are available with the default setup, we designed two 
other calibration approaches. In the former, a custom designed Cockroft-Walton accelerator feeding a 
proton beam with $E_p\approx 1\,MeV$ onto a target containing Boron. The reaction 
$B(p,\gamma \gamma )C$ delivers two simultaneous $\gamma$ with $E_\gamma^1=4.4\,\mathrm{MeV}$ and 
$E_\gamma^2=11.7\,\mathrm{MeV}$.
The $\gamma$s are detected in the TC and in the calorimeter providing a tool for 
bar intercalibration.

Another approach consists in delivering a $\pi^+$ beam on the target to produce $\pi^0$ through
charge exchange process. The Dalitz decay $\pi^0\rightarrow \gamma e^+e^-$ provides 
$\gamma$-$e^+$ pairs generated from the same $4d$ vertex. Using $t_\gamma$ as reference, the 
relative interbar offsets are 
calculated. The calorimeter-TC offset is also obtained with the same method.

The result of the interbar calibration at different times in Fig.\ref{fig3}a shows that the offsets
are remarkably stable. 

\begin{figure*}[!t]
\centerline{
\subfigure[Simulated $\mu\rightarrow e\gamma$ event]{\includegraphics[width=0.4\textwidth]{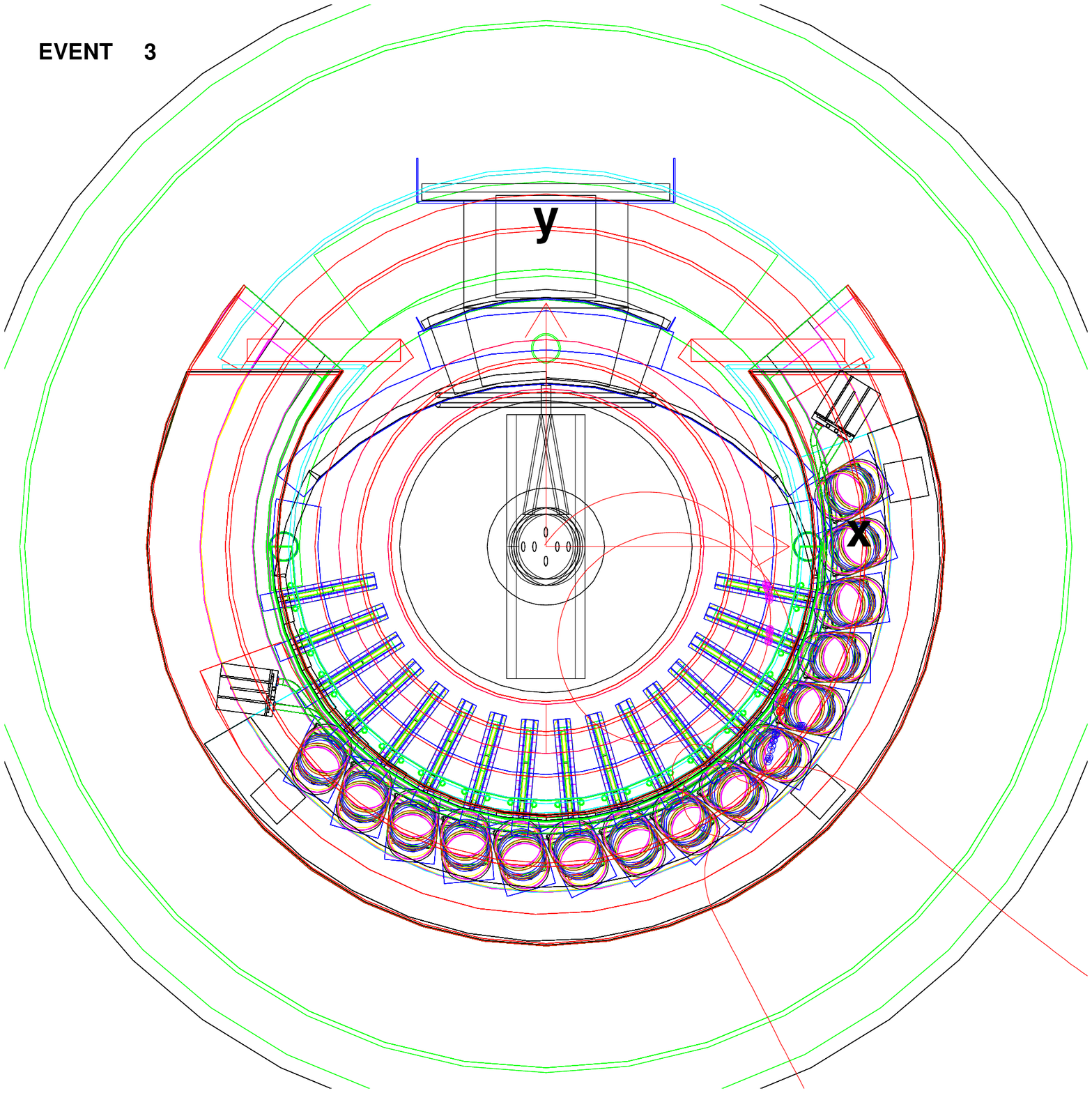}}
\hfil
\subfigure[Radiative Decay peak]{\includegraphics[width=0.4\textwidth]{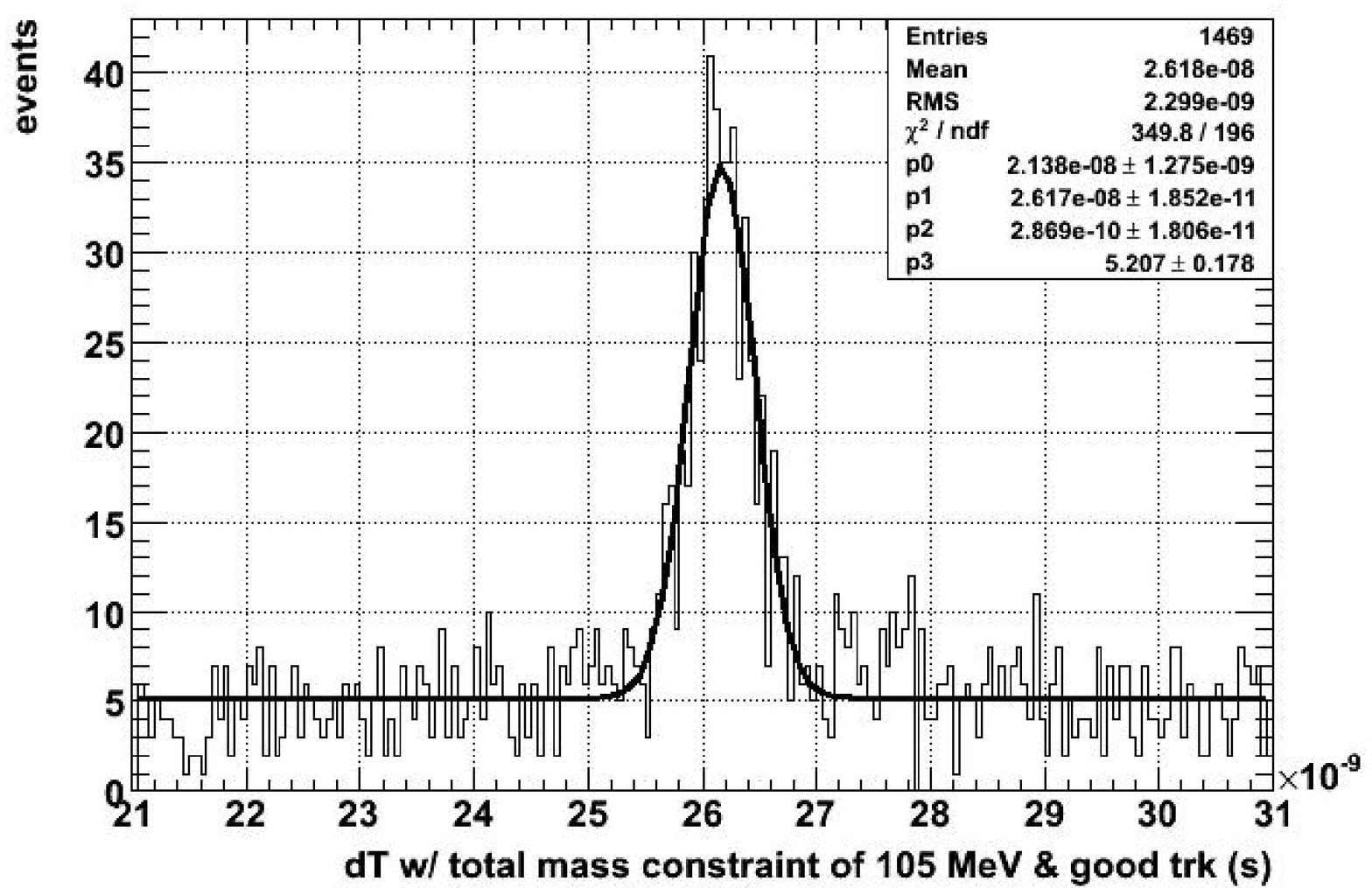}}
}
\caption{}
\label{fig12}
\end{figure*}

\begin{figure*}[!t]
\centerline{
\subfigure[Interbar offset calibration]{\includegraphics[scale=0.25,angle=-90]{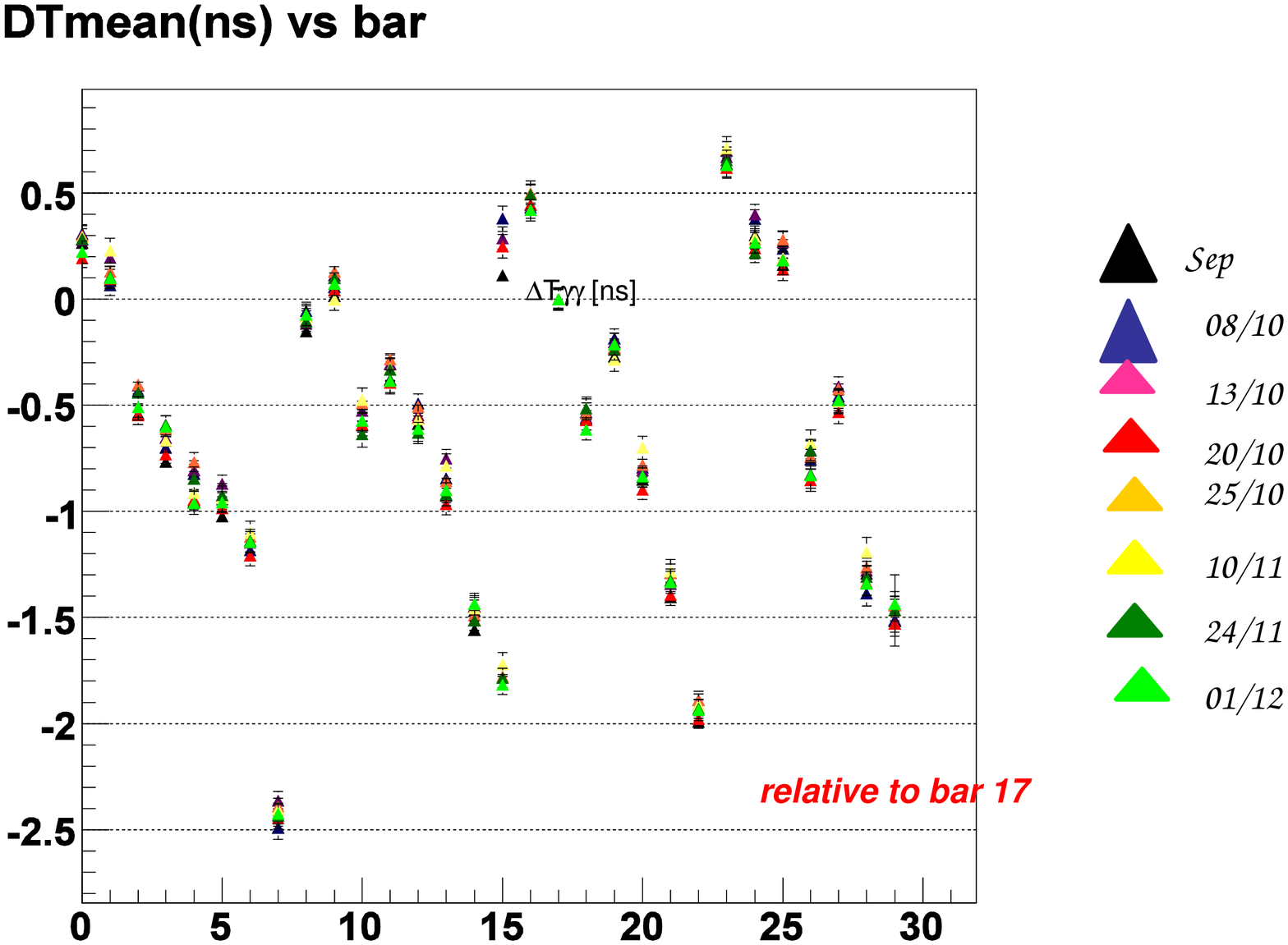}}
\hfil
\subfigure[DCH Track match with TC hit]{\includegraphics[scale=0.25,angle=-90]{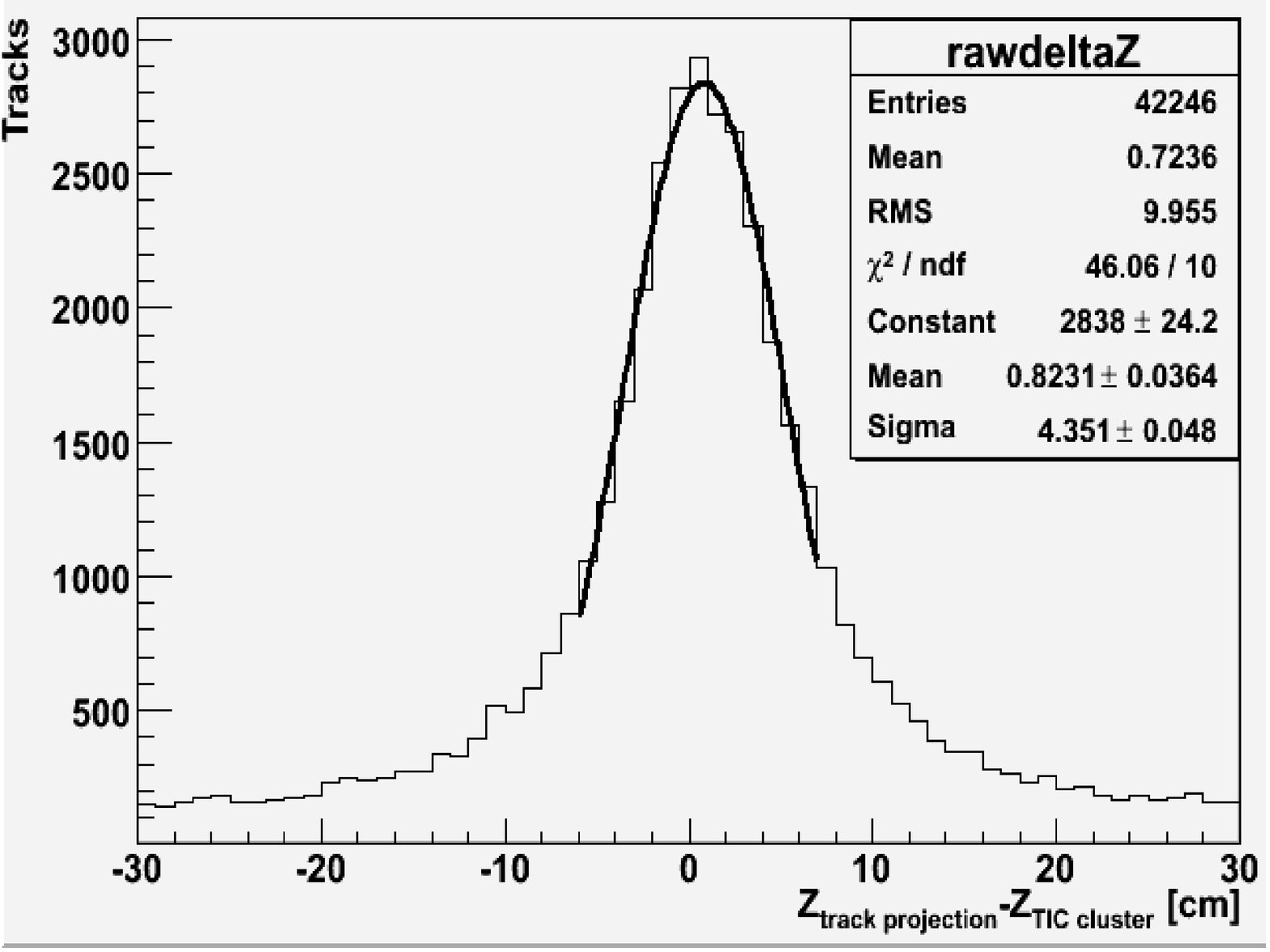}}
}
\caption{}
\label{fig3}
\end{figure*}

\section{The DCH-TC matching}

A measurement of $e^+$ requires both the DCH for the kinematic parameters and the TC for the time. 
The DCH track is extrapolated to the TC surface and matched with a TC hit, that provides
information about $\phi$ from the bar position, $z$ from the fiber position and from 
$T_{PMT0}^n-T_{PMT1}^n$. In Fig.\ref{fig3}b the DCH-TC match in $z$ is shown with a resolution 
$\approx 4\,cm$. 

\section{Radiative Decay detection}

One of the most critical tests of the feasibility of the experiment is its capability of detecting 
$e^+\gamma$ pairs emitted from the same space-time point, distinguishing them from the combinatorial 
background. These pairs are generated from $\mu^+$ Radiative Decays with a known Branching Ratio and its
detection is a proof of the good time resolution of the spectrometer. 
In Fig.\ref{fig12}b the Radiative Decay peak is clearly visible above the background.

\section{Conclusions}

The MEG spectrometer was commissioned during 2008. A hardware problem with the DCH limited its 
efficiency and momentum resolution. Nevertheless several aspects of the spectrometer functionality were 
successfully tested. 

The detection of RD events demonstrates the functionality of the apparatus.

\end{document}